\documentclass[reprint,nofootinbib,showpacs,showkeys]{revtex4-1}
\usepackage{palatino,url}
\usepackage{amsmath}
\usepackage{graphicx}
\usepackage{multirow}
\usepackage{bm}

\begin{document}

\title{CMB dipoles and other low-order multipoles in the quasispherical Szekeres model}
\date{\today}
\author{Robert G. Buckley}
\email{tcp385@my.utsa.edu}
\author{Eric M. Schlegel}
\email{eric.schlegel@utsa.edu}
\affiliation{University of Texas at San Antonio}


\begin{abstract}
Several authors have previously shown that Gpc-scale void based on the spherically symmetric LTB model can provide a good fit to certain cosmological data, including the SNIa data, but it is only consistent with the observed CMB dipole if we are located very close to the center, in violation of the Copernican principle. In this work we investigate the more general quasispherical Szekeres model, which does not include spherical symmetry, in order to determine whether this option may be less constricting. We find that the observer is still constrained to a small region, but it is not as geometrically ``special'' as the center of an LTB void. Furthermore, whereas the quadrupole and octupole near the center of an LTB void are necessarily small, certain Szekeres models can include a significant quadrupole while still being consistent with the observed dipole, hinting that Szekeres models may be able give an explanation for the observed quadrupole/octupole anomalies.
\end{abstract}

\maketitle

\section{\label{sec:intro}Introduction}

The current standard model of the universe includes a large, mysterious dark energy component, generally taken to be a cosmological constant, $\Lambda$. We know virtually nothing about this major part of the universe besides its magnitude, inferred primarily from the supernova luminosity-redshift relation \cite{Riess}. The standard $\Lambda$CDM model also provides a good fit to the CMB power spectrum \cite{Larson} and baryon acoustic oscillation data \cite{Percival}, but these all essentially measure the same thing: luminosity distances \cite{Durrer}. The cosmological constant, a key feature of the standard model of cosmology, hinges on the assumption that our interpretation of this one quantity is accurate. In recent years, a number of authors have suggested that an inhomogeneous universe model could provide an alternative explanation of these observations without requiring dark energy \cite{Buchert,Wiltshire,Alnes,Celerier,GBH,Clarkson}. When we see larger luminosity distances than expected, it could be due to a decrease in the expansion rate with distance, rather than an increase with time. Much work has gone into studying the Lema\^itre-Tolman-Bondi (LTB) model, an exact spherically symmetric solution to Einstein's equations \cite{Celerier2,GBH,GBH2,Romano,Zibin,Zibin2,Yoo,Biswas,Alnes2,Foreman}. This model is capable of matching any possible distance-redshift curve, without the need for any sort of dark energy.

This approach has a coincidence problem of its own. To fully explain the supernova luminosity-redshift data with a local void, it must be very large---at least several hundred Mpc in radius \cite{Bolejko2}. It is unrealistic to put us at the exact center of such a model, but if we are too far off-center, we would see a much larger CMB dipole than what we actually observe, since photons passing through the center of the void experience more of the higher expansion rate inside the void, and are also subject to a large-scale Rees-Sciama effect \cite{GBH2}. Alnes first estimated the relationship between the observer's position and the CMB dipole in \cite{Alnes}, and later calculated that this constrains us to a position within 15 Mpc of the void center for a 1500 Mpc-radius void \cite{Alnes2}. Foreman later calculated the constraint at 80 Mpc using a different model and somewhat different methods, still a small fraction of the total void radius \cite{Foreman}. This goes against the Copernican principle, which states that the Earth does not occupy a special place in the universe. Indeed, to claim that we are very close to the symmetry center of the universe would seem to be a step backwards towards the geocentric worldviews of antiquity. Still, the Copernican principle is an assumption, and though it has gained support from recently proposed tests \cite{Valkenburg,ClarksonCopernican,BolejkoCopernican,Caldwell,Zhang}, it is not yet rigorously established by observations. It should not be dismissed lightly, but there is still room to consider alternative models.

Even if the observer is lucky enough to be in this small low-dipole region, the high dipoles seen by hypothetical observers farther from the center poses a problem due to the kinetic Sunyaev-Zel'dovich (kSZ) effect. Free electrons scatter CMB photons towards the observer, and if those electrons see a large dipole along the line of sight, this will affect the observed spectrum. This creates an additional contribution to the CMB power spectrum at small angular scales, tracing the anisotropy of the projected free electron surface density \cite{GBH2,Zibin,Zhang,Yoo2}. This effect was first studied in relation to clusters in LTB void models in \cite{GBH2}, then estimated for intra-cluster gas in \cite{Zhang}, and finally shown by Moss and Zibin to rule out most LTB models without dark energy in \cite{Zibin}. 

Furthermore, due to the symmetry, CMB anisotropies in the next several multipoles beyond the dipole receive very little contribution from the inhomogeneity for observers near the center. This is disappointing, because one of the most significant anomalies seen in the WMAP CMB data is the improbable alignment of the quadrupole and octupole, first pointed out by Tegmark \cite{Tegmark}. The preferred axes of these two multipoles lie within $1\,^{\circ}$, due to no specific feature, and there is currently no model to explain this \cite{Bennett}. One might imagine that a very-large-scale inhomogeneity of the sort proposed to explain the distance-redshift curve could explain these large-scale anomalies, but Alnes \cite{Alnes2} found that, for observers in the region allowed by the dipole, such a void produces only a very small quadrupole and octupole, insignificant compared to what is found in the WMAP.

The LTB model, though more general than the homogeneous and isotropic Friedmann-Lema\^itre-Robertson-Walker (FLRW) model, is still a simplification. It can be considered a smoothing of the inhomogeneities over the angular variables, keeping only variation with respect to the radial direction. The next step towards a general inhomogeneous universe model is the Szekeres class of models introduced in \cite{Szekeres}. These models, though still not completely general, possess no inherent symmetries. The most relevant subclass, the quasispherical Szekeres model, can be pictured like an LTB model, but with the spherical shells shifted around relative to each other. The notion of a ``center'' thus becomes somewhat unclear; outer shells are not generally centered at the coordinate origin. Ishak {\itshape et al.} and others have argued that this property gives these models an advantage over LTB models with regards to the Copernican principle \cite{Ishak,BolejkoCoarse}. If there is no single unique center, our position may not be so special after all.

Nevertheless, we must still satisfy the requirement that the CMB dipole seen at the observer is not unacceptably large compared to observations. We must then ask, in what region of a Szekeres model of the kind Bolejko proposes would an observer see a suitably small dipole? How does the volume of this region compare to that of the corresponding LTB model? If the region is still small, it would seem that even in the Szekeres model we must reside in a special location---the place where the observed dipole is small---even if it is not the ``center''. This provides a more quantitative test of the model's compliance with the Copernican principle. 

Once we have located the low-dipole region, we can also investigate other properties this region can have, such as the CMB quadrupole and octupole. This may show further advantages over LTB---in LTB, the low-dipole region has a inhomogeneity-induced quadrupole and octupole too small to explain the anomalous alignment seen in the real CMB \cite{Alnes2}, but we should not expect Szekeres to be so limited. Studying the dipoles across the void will also provide hints on whether Szekeres models suffer the same constraints from the kSZ effect as LTB models.

In summary, the LTB model has four shortcomings related to its symmetry, the Copernican principle, and the CMB, against which we wish to test the Szekeres model:
\begin{itemize}
\item Quantitatively, there is only a small region in an LTB universe model in which an observer would see a CMB dipole consistent with observations. This fine-tuning requirement violates the Copernican principle, which implies that any location should be equally valid.
\item Qualitatively, the LTB model further violates the Copernican principle because this ``allowed'' region is geometrically special.
\item Due to the symmetry, CMB anisotropies in the next several multipoles beyond the dipole receive very little contribution from the inhomogeneity, so the LTB model offers no explanation for the observed anomalies in the quadrupole and octupole.
\item The kSZ effect at $l \simeq 2000\text{--}3000$ is too strong to be reconciled with observations \cite{Zibin}.
\end{itemize}

The rest of this paper is organized as follows. In Section \ref{sec:model}, we present the equations governing the quasispherical Szekeres class of models and give the function definitions used to describe the test models we will use. Section \ref{sec:methods} describes the methods we will use to perform our dipole calculations. Section \ref{sec:theory} gives a brief theoretical discussion of how exactly a CMB dipole arises in such models. We present and discuss our results in Section \ref{sec:results}, and give our conclusions in Section \ref{sec:conclusion}.

\section{\label{sec:model}Model Definitions}

\subsection{The Szekeres model}

The Szekeres model is a generalization of the LTB model. It, too, contains only a comoving, irrotational, pressureless dust. The Szekeres model, however, in general has no symmetry; there are no Killing vectors, except in special cases \cite{Bonnor}.

The quasispherical Szekeres model is described by the metric
\begin{equation}
\label{eq:metric}
\mathrm{d}s^2 = -\mathrm{d}t^2 + \frac{(\Phi' - \Phi\frac{E'}{E})^2}{1-k}\mathrm{d}r^2 + \frac{\Phi^2}{E^2}(\mathrm{d}x^2 + \mathrm{d}y^2).
\end{equation}
Here, $\Phi = \Phi(t,r)$ is the areal radius of the spherical shell labeled by $r$ at time $t$, $k = k(r)$ is an arbitrary function determining curvature, and $E = E(r,x,y)$ describes the departure from LTB. A prime denotes a partial derivative with respect to $r$. The function $E(r,x,y)$ is defined in terms of three arbitrary functions of $r$ as
\begin{equation}
E(r,x,y) = \frac{[x-P(r)]^2 + [y-Q(r)]^2 + S(r)^2}{2S(r)}.
\end{equation}
As with the LTB model, this model consists of a series of spherical shells labeled by the coordinate $r$. The coordinates on the shell, $x$ and $y$, relate to the standard $\theta$ and $\phi$ by a stereographic projection, as we will explain in the next subsection. Unlike LTB, these shells are not concentric, nor is matter distributed evenly across a given shell. The functions $P(r)$, $Q(r)$, and $S(r)$ have three effects on the model:

\begin{itemize}
\item They displace the centers of the shell $r + \delta r$ relative to the shell $r$ by $\delta r \:\Phi P'/S$ in the direction $(\theta , \phi )=(\pi/2, 0)$, by $\delta r \:\Phi Q'/S$ in the direction $(\pi/2, \pi/2)$, and by $\delta r \:\Phi S'/S$ in the direction $(0, 0)$;
\item They rotate the shells by $\delta r \: P'/S$ about the axis $(\pi/2, -\pi/2)$ and by $\delta r \: Q'/S$ about the axis $(\pi/2, 0)$;
\item They redistribute the matter on each shell in the shape of a dipole along the direction of shifting.
\end{itemize}
If $P'$, $Q'$, and $S'$ all vanish, the model reduces to LTB.

The Einstein equations describe the evolution of the model in terms of its matter distribution and curvature:
\begin{equation}
\label{eq:rdot}
\frac{\dot{\Phi}(t,r)^2}{c^2} = \frac{2M(r)}{\Phi(t,r)} - k(r) + \frac{1}{3}\Lambda\Phi(t,r)^2,
\end{equation}
where an overdot indicates $\partial/\partial t$, $\Lambda$ is a possible cosmological constant, and the function $M(r)$ is related to the density by
\begin{equation}
\label{eq:density}
4\pi\frac{G}{c^2}\rho(t,r,x,y) = \frac{M'(r) - 3M(r)\frac{E'(r,x,y)}{E(r,x,y)}}{\Phi(t,r)^2\left[\Phi'(t,r) - \Phi(t,r)\frac{E'(r,x,y)}{E(r,x,y)}\right]}.
\end{equation}
By integrating (\ref{eq:rdot}), we reveal another free function:
\begin{equation}
\label{eq:bangtime}
t - t_B(r) = \int_0^\Phi{\frac{dR}{\sqrt{2M/R - k + \Lambda R^2 /3}}}.
\end{equation}
The function $t_B(r)$ is called the ``bang-time function'', because it denotes the time at which the shell labeled by $r$ emerges from the big bang singularity. It is associated with decaying modes. For the remainder of this paper, we will set $\Lambda = 0$.

There is a gauge freedom in the choice of the $r$ coordinate, since the model is covariant under transformations of the form $\tilde{r} = f(r)$. For instance, we could choose $r$ so that $\Phi(t_0,r) = r$, where $t_0$ is the present time. This effectively fixes $t_B(r)$ in terms of $k(r)$ and $M(r)$ through Eq.\ (\ref{eq:bangtime}). This leaves five free functions of $r$ to define the model: $M$, $k$, $S$, $P$, and $Q$.

\subsection{Spherical coordinates}

We can bring the coordinates to a more familiar form with a simple transformation:
\begin{subequations}
\label{eq:sphericaltransform}
\begin{equation}
x - P = S \cot\left(\frac{\theta}{2}\right) \cos \phi,
\end{equation}
\begin{equation}
y - Q = S \cot\left(\frac{\theta}{2}\right) \sin \phi.
\end{equation}
\end{subequations}
In these coordinates, the metric is significantly more complicated and no longer diagonal, but for some applications they provide greater clarity. For instance, we can write
\begin{equation}
\frac{E'}{E} = -\frac{S' \cos \theta + (P' \cos \phi + Q' \sin \phi)\sin \theta}{S}.
\end{equation}
This makes it clear that $P$ defines anisotropy in the direction $(\theta = \pi/2, \phi = 0)$, $Q$ in the direction $(\theta = \pi/2, \phi = \pi/2)$, and $S$ in the direction $(\theta = 0)$---what we would call ``$x$'', ``$y$'', and ``$z$'' in pseudo-Cartesian coordinates. Note that the positive ``$z$'' axis has the Szekeres $x$ and $y$ coordinates diverge; we will have to steer clear of this region to avoid problems with our numerical calculations. There is nothing physically special about this region (except in the case of $P' = Q' = 0$ and $S' \neq 0$, but even then a simple coordinate transformation can switch the positive and negative ``$z$'' directions), so systematically avoiding this region should not significantly affect our analysis.

\subsection{Test models}

We will construct a set of Szekeres test models by starting with one base LTB model and adding several different Szekeres functions to it, each one resulting in a different Szekeres model.

\subsubsection{Base LTB model}

For a base LTB model, we use a constrained Garcia-Bellido Haugb\o lle (GBH) model \cite{GBH}. This model describes a large void with a homogeneous Big Bang (that is, with $t_B(r) = const$), with a density profile defined in terms of a radially-dependent matter density parameter, defined as
\begin{align}
\Omega_M (r) = \Omega_{out} + (\Omega_{in} - \Omega_{out})\left\{\frac{1 - \tanh{[(r-r_0)/2\Delta r]}}{1 + \tanh{(r_0 / 2\Delta r)}}\right\},
\end{align}
and a radially-dependent present expansion rate $H_0 (r)$, defined to ensure $t_B(r) = const$.\footnote{A similar model was introduced by Alnes {\itshape et al.} in \cite{Alnes}. However, we followed Garcia-Bellido and Haugb\o lle's presentation because we found it made certain aspects more intuitive, such as their method of ensuring $t_B(r) = const$.} Since our model does not include dark energy, we need $\Omega_{out} = 1$ in order to ensure asymptotic flatness far from the void. Our choices for the other parameters, $\Omega_{in}$ (the matter density at the center of the void), $r_0$ (the characteristic size of the void), $\Delta r$ (the sharpness of the void wall), $t_0$ (the present age of the universe), and $H_0$ (the local Hubble constant at the center of the void) are given in Table \ref{tab:modelparameters}. This is similar to the best-fit model found in \cite{GBH}, which was selected by combining SNIa distance-redshift data, baryon acoustic oscillation (BAO) data, and the scale of the first peak in the CMB power spectrum. We define our $r$ coordinate so that $\Phi(t_0,r) = r$ in units of Mpc.

\subsubsection{Szekeres functions}

For our test models, we desired something simple enough to be readily analyzable, yet without symmetries which could hide more general effects. We chose to set $S(r) = 1$ and $Q(r) = 0$. This leaves only one function to work with, $P(r)$, yet does not result in axial symmetry (though there is a discrete bilateral symmetry). We constructed our $P(r)$ functions by a method inspired by \cite{BolejkoCoarse}. First, we define a function
\begin{align}
d(r,r_i,r_f) = (1+r)^{-0.99} e^{-0.0003r} \frac{(r - r_i)^2(r_f - r)^2}{(r_f/2 - r_i/2)^4}.
\end{align}
Then, we define $P(r)$ as a piecewise function,
\begin{align}
P(r) = C \left\{ 
  \begin{array}{l l}
  	0 & \quad r < r_i \\
  	\int_{r_i}^r {d(\tilde{r},r_i,r_f)\mathrm{d}\tilde{r}} & \quad r_i < r < r_f \\
  	\int_{r_i}^{r_f} {d(\tilde{r},r_i,r_f)\mathrm{d}\tilde{r}} & \quad r > r_f
  \end{array} \right.
\end{align}
This general form allows us to put a Szekeres anisotropy of any strength we wish over any $r$ range we wish, depending on the constant parameters $C$, $r_i$, and $r_f$. As long as $C < 1$, there is no shell crossing at the present time, and the last factor in the definition of $d(r,r_i,r_f)$ ensures that $P(r)$ is continuous up to the second derivative, avoiding possible numerical issues.

In this manner, we construct six test models, with parameters given in Table \ref{tab:modelparameters}. In model 1, the $P$ function is moderately strong and extends from the origin to one fourth the void radius. In model 2, the $P$ function is weaker, but covers a broader range, and does not begin until one fourth the void radius. This allows us to separate the radial dependence of the dipole from local effects of the Szekeres function. The third model's $P$ function has only a relatively narrow spike, allowing us to examine the effects of an isolated segment of Szekeres anisotropy from locations in the interior, exterior, and middle of the anisotropic shells. These three models will be the focus of our investigation, but we will also examine three more: model 4, which is like 1 but with a stronger $P$ function, model 5, again like 1 but with broader range, and model 6, like 3 but with the spike at a much higher $r$ value, to compare the effects of distant anisotropies and nearby ones. 

\begin{table}[tbp]
	\begin{center}
		\begin{tabular*}{8.6cm}{@{\extracolsep{\fill}}ccccccr@{.}lrr}
			\hline \hline
				\multicolumn{5}{c}{Base LTB Model Parameters} & & \multicolumn{4}{c}{Szekeres Parameters} \\
				\cline{1-5} \cline{7-10}
				 $\Omega_{in}$	&	$r_0$ & $\Delta r$ & $t_0$ & $H_0$ & Model & \multicolumn{2}{c}{$C$} & \multicolumn{1}{c}{\ \ $r_i$} &  \multicolumn{1}{c}{\ \ $r_f$} \\
				& Mpc & Mpc & Gyr & $\frac{\text{km}}{\text{s Mpc}}$ &  & \multicolumn{2}{c}{} & \multicolumn{1}{c}{\ \ Mpc} & \multicolumn{1}{c}{\ \ Mpc} \\
				\hline
				&	&	&	&	&	1 & 0&630 & \quad 0 		&  \quad 575 	 \\ 
				& & & & & 2 & 0&315 & \quad 575 	&  \quad 2875 	 \\
				\multirow{2}{*}{0.13} & \multirow{2}{*}{\ 2300} & \multirow{2}{*}{\ 620} & \multirow{2}{*}{\ 15.3} & \multirow{2}{*}{64} &
									3 & 0&945 & \quad 100 	&  \quad 300 	 \\
				& & & & & 4 & 0&945 & \quad 0 		&  \quad 575 	 \\
				& & & & & 5 & 0&630 & \quad 0 		&  \quad 1150 	 \\
				& & & & & 6 & 0&945 & \quad 1900 	&  \quad 2100 	 \\
				\hline \hline
		\end{tabular*}
	\caption{The parameters used to define our test models. The same base LTB model parameters apply to all 6 models. Quantities in Mpc refer to area distances of shells at the present time.}
	\label{tab:modelparameters}
	\end{center}
\end{table}

\begin{figure}[tbp]
	\begin{center}
(a)\\	\includegraphics[width=8.0cm]{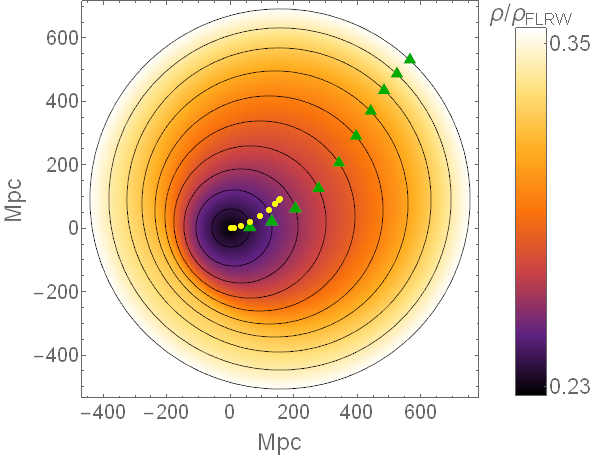} \\
(b)\\ \includegraphics[width=8.0cm]{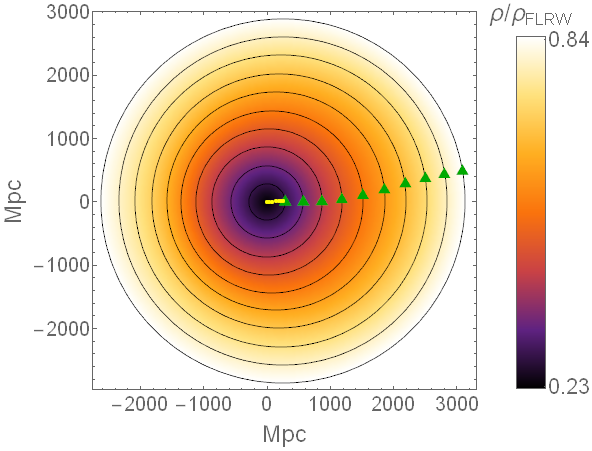} \\
(c)\\ \includegraphics[width=8.0cm]{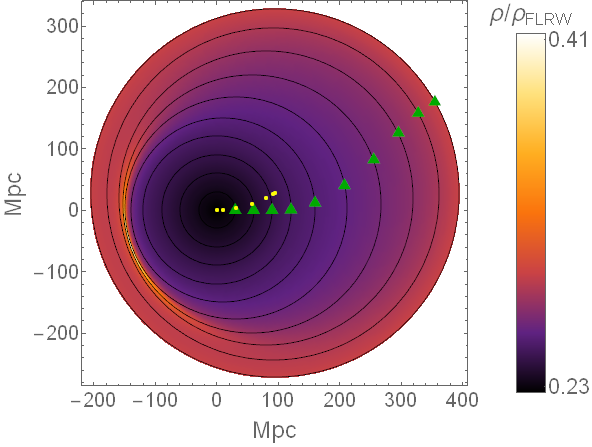}
	\caption{Density plots of (a) model 1, (b) model 2, and (c) model 3, covering a two-dimensional cross-section corresponding with the symmetry plane. The plotting range for each model is chosen to cover the Szekeres anisotropies, and the color scale is adjusted for each plot to maximize the visual contrast. Densities on the scale are written as fractions of $\rho_{FLRW}$, the density of the background FLRW model, which the test models approach asymptotically at very high $r$. Black circles show shells of constant $r$. The green triangles on each marked shell show the direction of shell shifting, and yellow dots show the geometric centers of the shells. (Color online)}
	\label{fig:densityplots}
	\end{center}
\end{figure}

Figure \ref{fig:densityplots} shows two-dimensional cross-sections of the density distributions of each of the first three models. These are {\bfseries not} intended to be realistic models. Their purpose is to provide insight into the observational effects that can arise from a Szekeres-type anisotropy and to establish a baseline from which we can extrapolate to more general cases. A more realistic model would require that an observer in the region allowed by the dipole would also see a luminosity distance-redshift curve with directional variation within the constraints set by supernova observations, as well as consistency with baryon acoustic oscillations, galaxy age data, and other such observations (all now direction-dependent due to the lack of perfect isotropy), all while containing structures in some manner consistent with the shape and statistics of observed large-scale structure.

\section{\label{sec:methods}Methods}

\subsection{Calculating the dipole}

We can calculate the observed CMB temperature at any point in the sky by generating a null geodesic from the observer backwards in time to the last scattering surface (LSS), with the initial tangent vector at an angle corresponding to the point in the sky in question. We do this by integrating the geodesic equations,
\begin{align}
\label{eq:geodesicequation}
\frac{\mathrm{d}^2 x^\mu}{\mathrm{d}\lambda ^2} + \Gamma^\mu_{\alpha \beta} \frac{\mathrm{d}x^\alpha}{\mathrm{d}\lambda} \frac{\mathrm{d}x^\beta}{\mathrm{d}\lambda} = 0,
\end{align}
where $\Gamma$ is the Christoffel symbol and $\lambda$ is an affine parameter along the photon path. The exact forms of the geodesic equations in the Szekeres model can be found in \cite{Nwankwo} and \cite{KrasinskiRedshiftProp}, and are reproduced in Appendix \ref{app:geoeq}, along with a brief discussion on calculating redshift, which determines the CMB temperature seen along any line of sight.

To rigorously calculate the observed dipole, we would have to generate geodesics going in every direction to see the CMB temperature over the whole sky, and find the dipole by calculating the $a_{1m}$ coefficients of the spherical harmonic expansion,
\begin{align}
\label{eq:fulldipole}
D &= \sqrt{\sum_{m = -1}^1 {\left|a_{1m}\right|^2}} \\
a_{1m} &= \int_0^{2\pi} {\int_0^\pi {\frac{\Delta T}{T} Y_{1m}(\theta,\phi) \sin{\theta} \, \mathrm{d}\theta \, \mathrm{d} \phi }}
\end{align}
Since we desire a map of the observed dipole over the whole space of the model, this would be computationally expensive. A much faster method is possible if we can assume the CMB anisotropies are dominated by the dipole term. This is the case near the center of an LTB model \cite{Foreman,Alnes2}; we will have to check whether this still holds in our Szekeres models.

The method we will use is an extension of that used in \cite{Foreman}. We will generate three spatially orthogonal pairs of null geodesics backwards in time from the observer, with the geodesics in each pair propagating in opposite spatial directions. A basic illustration is shown in Fig.\ \ref{fig:6geo}, and the precise methods used to choose directions are described in Appendix \ref{app:choosingdirections}. 

\begin{figure}[tbp]
	\begin{center}
	\includegraphics{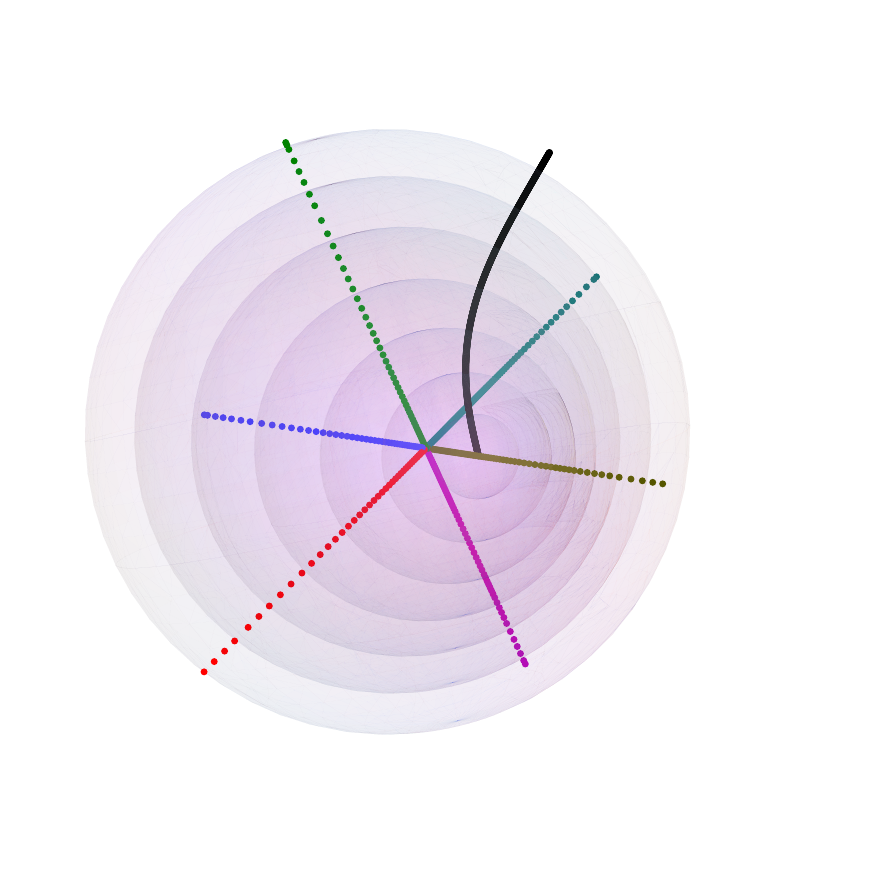} 
	\caption{The six geodesics used to calculate the dipole at a particular point in model 5. Each geodesic is shown in a different color, and each dot represents one step in the numerical integration. The black line indicates where $x$ and $y$ diverge to $\pm \infty$. (Color online)}
	\label{fig:6geo}
	\end{center}
\end{figure}

Assuming the temperature of the LSS is uniform, the CMB temperature measured at any point in the sky is found from the redshift of the geodesic in that direction by
\begin{equation}
\label{eq:geoT}
T = \frac{T_{*}}{1+z_{*}},
\end{equation}
where asterisks mark quantities at the LSS. We further assume that the intersections of the geodesics with the LSS occurs at an equal time $t_{*}$ in the synchronous gauge, regardless of direction of propagation.\footnote{This is not strictly accurate, as the void in our test models approaches FLRW only asymptotically, without a compensating overdensity; since geodesics going in different directions from a non-central observer reach different radial distances, the LSS may occur at slightly different times for each. However, in our calculations, all of the geodesics reach distances where the density approaches FLRW closely enough that such differences are insignificant, as further verified in the next section.} The temperature difference in each pair of geodesics can be treated as a component of a vector. The magnitude of this vector gives the total dipole. By dividing by the mean temperature, we get the apparent dipole velocity,
\begin{equation}
\label{eq:6dipole}
v = \frac{\sqrt{\left(T_1 - T_2\right)^2 + \left(T_3 - T_4\right)^2 + \left(T_5 - T_6\right)^2}}{\frac{1}{3}\left(T_1 + T_2 + T_3 + T_4 + T_5 + T_6\right)},
\end{equation}
and the dipole magnitude is given by $D = \sqrt{\frac{4\pi}{3}}v$. For a derivation confirming that $v$ as calculated above is indeed the correct dipole, and a brief assessment of the error caused by higher order multipoles, refer to Appendix \ref{app:6geo}.

In practice, we do not need to integrate the geodesics all the way back to the LSS. At large radii, our models asymptotically approach FLRW. If we integrate to a sufficiently early time $t_1$, all of the geodesics will be far enough outside of the inhomogeneity that from then on the redshifts evolve nearly exactly as in FLRW. Then we can write
\begin{equation}
\label{eq:intermediateT}
T = \frac{T_{*}}{1+z(t_1)} \times \frac{a(t_1)}{a(t_{*})},
\end{equation}
where $a(t)$ is the scale factor of the FLRW background. The factors $\frac{a(t_1)}{a(t_{*})}$ in the numerator and denominator of (\ref{eq:6dipole}) cancel out, and can therefore be ignored. Likewise, we do not need to assume any particular value for $T_{*}$, since it does not affect the final result in (\ref{eq:6dipole}).

To check that the result is the true dipole, and has not been overly contaminated by higher multipoles, we can repeat the process with a different set of orthogonal geodesics, and compare the results. Using this method, we estimate that the relative error in our data due to this effect is on the level of $10^{-3}$ or less.

\subsection{Higher order multipoles}

We have seen how to calculate the CMB dipoles generated by the inhomogeneities, but this is not the only effect the inhomogeneities have on the CMB. The inhomogeneities leave higher-order multipole imprints on the CMB as well. To analyze the extent of the inhomogeneity-induced spherical harmonics, we adopt a procedure similar to that used to calculate the dipoles, but with many more geodesics, propagating in evenly spaced directions across the entire sky. We will use a spacing of 4 degrees, for a total of 2534 data points for each location we test. 

To obtain the strength of a given multipole, we calculate the $a_{lm}$ coefficients by numerical integration (limited by the resolution of the data)\footnote{In practice, we remove each multipole (starting with the monopole) from the data before calculating the next, to avoid spurious results from integration error.}:
\begin{align}
\label{eq:alm}
a_{lm} \approx \sum_{n=1}^{2534} {T(\theta_n , \phi_n) Y^{*}_{lm}(\theta_n , \phi_n) \sin{\theta_n} \: \delta \theta \: \delta \phi}
\end{align}
We pixelise the sphere in a rectangular manner, with rows of points of constant $\theta$ and a uniform spacing between rows of $\delta \theta = 4\,^{\circ}$. Within each row, $\delta \phi$ varies to fill the circle\footnote{This simplistic pixelisation scheme is prone to certain systematic errors in the calculations, but tests suggest that in the present work these errors are on the order of 1 $\mu$K or less, small enough to be ignored. We used this scheme because it was simple to implement in Mathematica, but for future work we intend to use the HEALPix scheme, which is less straightforward to connect with our Mathematica code but ultimately more reliable.}:
\begin{align}
\delta \phi = \frac{180\,^{\circ}}{\lfloor 180 \sin{\theta}/4 \rfloor}.
\end{align}

\section{\label{sec:theory}Origin of the dipole}

In the LTB model, one can understand the CMB dipole in terms of the Rees-Sciama effect \cite{GBH2}. Since the void is in the nonlinear regime, its density contrast grows faster than the scale of the universe, causing the gravitational potential well to deepen over time. CMB photons passing through the void lose energy because the well they climb out of is deeper than the well they fell into. An off-center observer will therefore see that photons which pass through the center of the void are redshifted more than those coming from the opposite direction.

Another way to understand the dipole is by directly looking at two null geodesics, one passing through the center of the void (ingoing) and the other extending radially in the opposite direction (outgoing). If the observer is near the center, the ingoing geodesic crosses the center and returns to the original shell without picking up much redshift. The two geodesics then both propagate outwards (and backwards in time), but since the ingoing geodesic took a nonzero amount of time to cross the center (linearly proportional to the initial shell radius), they cross each shell at slightly different times throughout the journey.

We identify three potential ways in which the Szekeres functions can influence photon redshifts, and therefore the dipole:
\begin{itemize}
\item By directly affecting the longitudinal expansion rate. 
\begin{align}
\theta_l &= \frac{\dot{\Phi}' - \dot{\Phi}E'/E}{\Phi' - \Phi E'/E}
\end{align}
In a void model, this means that the expansion rate is slower where shells are pressed together, and faster where they are stretched apart. This induces greater photon redshift on the stretched side and lesser redshift on the compressed side.

\item By altering the times at which photons pass through shells. Looking backwards in time from the observer, photons traveling along the direction of shell shifting must travel a greater distance to reach the outer shells than for an observer at the same coordinates in the corresponding LTB model, thus reaching them at an earlier time; conversely, traveling in the opposite direction appears faster. Even when the Szekeres functions do not extend to high radii, this effect causes the observer to see the outer shells as though looking from a shifted position.

\item By influencing the total distance from the void the photons reach when they hit the surface of last scattering. This is related to the second case, but only applies when the model does not approach FLRW sufficiently quickly.
\end{itemize}

The third contribution is undesirable, since it explicitly violates the assumption of a statistically uniform surface of last scattering. We have confirmed that it does not play a significant role in our model by comparing the difference in the change in redshifts for a typical observer's geodesics between the times $t_0 / 600$ and $t_0 / 20000$. We find that they differ by less than 0.1\%. The dipoles found using these two times as ending times also differ by less than 0.1\%. We can therefore be confident that the dipole is not greatly influenced by effects near the surface of last scattering.

\section{\label{sec:results}Results and discussion}

For each model, we choose several $r$ values, and for each of these we calculate the CMB dipoles seen by observers at evenly spaced locations covering the sphere.

\subsection{Fitting function}

\begin{figure}[tbp]
\begin{center}
	(a) \qquad \qquad \qquad \qquad \qquad \qquad (b) \\
\includegraphics[width=4.25cm]{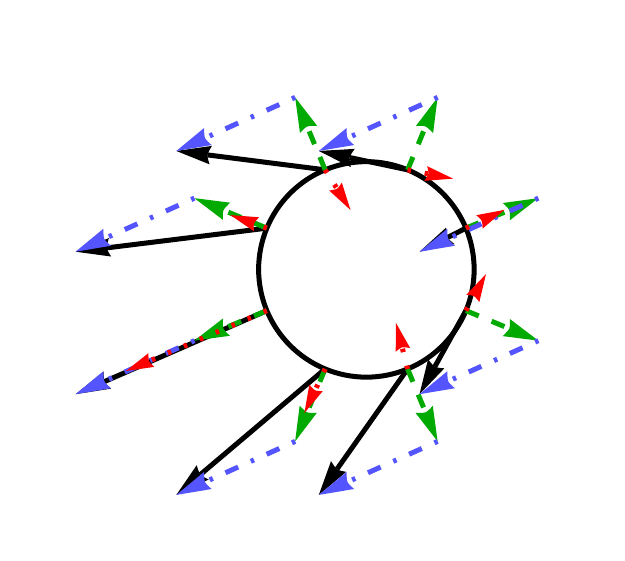}   
\includegraphics[width=4.25cm]{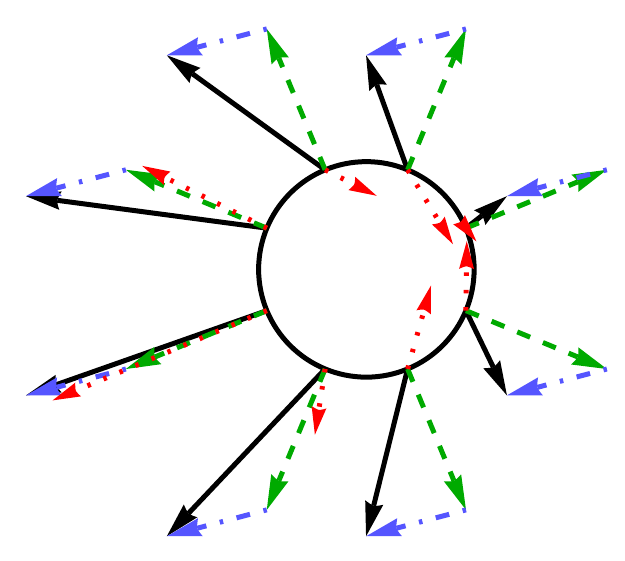} \\
(c) \qquad \qquad \qquad \qquad \qquad \qquad (d) \\
\includegraphics[width=4.25cm]{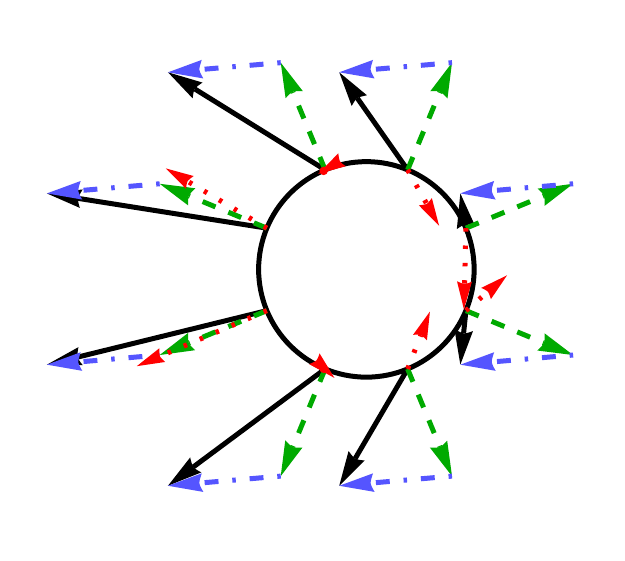}  
\includegraphics[width=4.25cm]{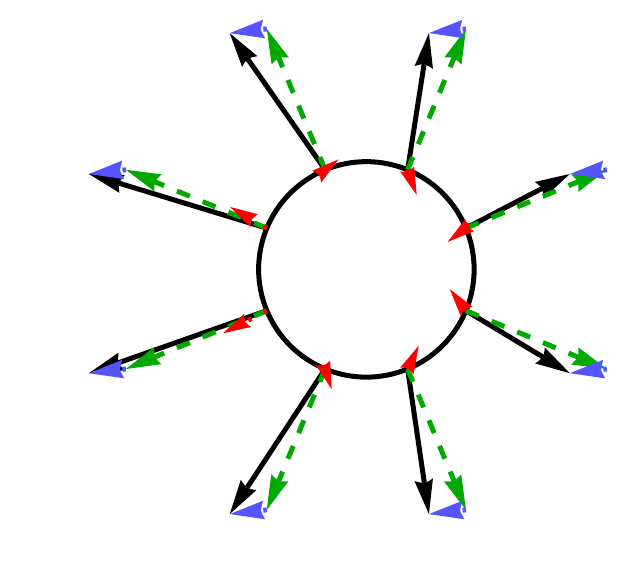}	
\end{center}
\caption{Magnitudes and directions of dipoles in a two-dimensional cross-section corresponding to the symmetry plane. Solid black arrows represent the total dipoles from numerical calculations, green dashed arrows are the LTB dipole, and blue dot-dashed arrows are the Szekeres component of the dipole. Smaller red dotted arrows are the fitting errors---the data minus the fit---magnified by a factor of 2000. (a): model 1, $r$ = 100 Mpc; (b): model 1, $r$ = 200 Mpc; (c): model 2, $r$ = 150 Mpc; (d): model 3, $r$ = 200 Mpc. (Color online) }
	\label{fig:dipolevectordecomp}
\end{figure}

We have found that the dipoles on each shell of constant $r$ can be well approximated by a simple function of three parameters:
\begin{align}
D(r,\theta, \phi) = a(r) \bm{\hat{r}}(\theta, \phi) + b(r)\left[\cos{\theta_0(r)} \bm{\hat{k}} - \sin{\theta_0(r)} \bm{\hat{i}}\right],
\end{align}
where $\bm{\hat{r}}(\theta, \phi)$ is the radial unit vector (normal to the shell), and $\bm{\hat{i}}$ and $\bm{\hat{k}}$ are unit vectors in the directions $(\pi/2, 0)$ and $(0,0)$, respectively. This describes the sum of two vectors, one of magnitude $a$ and radial direction, and one of magnitude $b$ and constant direction $(\theta_0, \pi)$. The magnitude of the former corresponds very closely to that of the dipole seen in the corresponding LTB model (i.e.\ a model with $S(r)$, $P(r)$, and $Q(r)$ set to constant values, but otherwise unchanged). The other vector can be thought of as a ``Szekeres dipole,'' as it is the result of the Szekeres $S(r)$, $P(r)$, and $Q(r)$ functions. A few examples are shown in Fig.\ \ref{fig:dipolevectordecomp}. This simple function is able to fit the data to within $0.1$ mK in all cases.

\begin{figure}[tbp]
	\begin{center}
	(a) \\ \includegraphics[width=8.5cm]{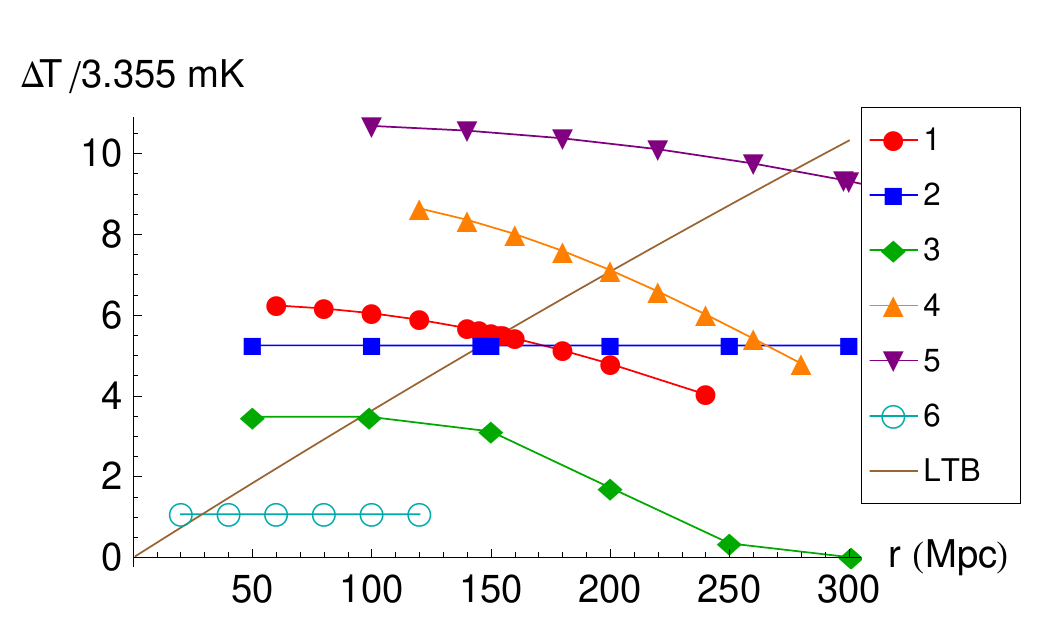} \\
	(b) \\ \includegraphics[width=8.5cm]{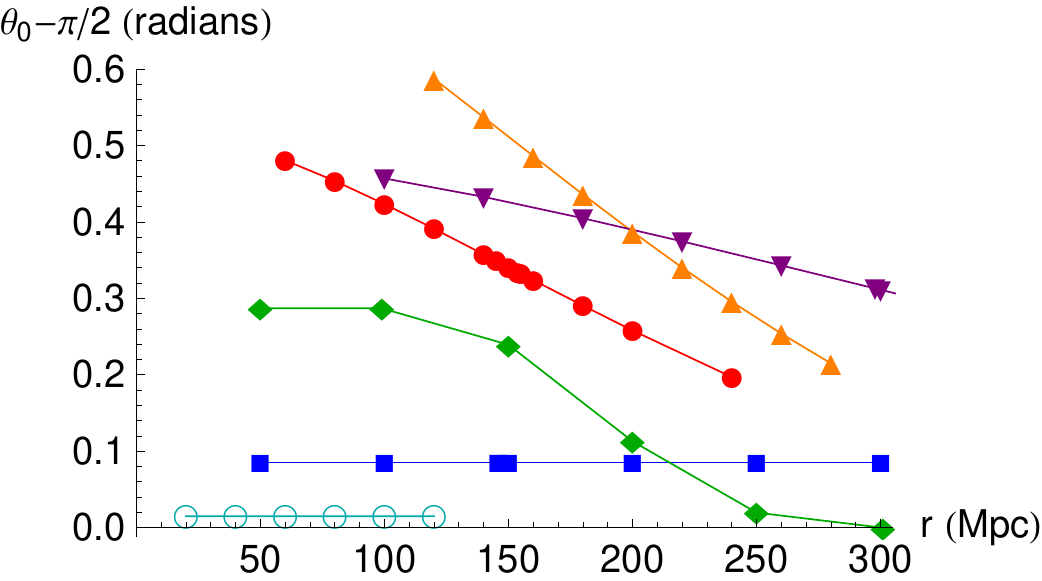}	
	\caption{(a): magnitudes of the Szekeres dipoles at various $r$ values in each of the three models, and the LTB dipoles in brown, all as a factor of the actual observed dipole. (b): $\theta_0$ for the same dipoles. (Color online)}
	\label{fig:szekdipoles}
	\end{center}
\end{figure}

Though the Szekeres dipole appears to be nearly constant on a given shell, its magnitude and direction do change as we move between shells. Figure \ref{fig:szekdipoles} shows the radial dependence for each of the six models. A few key features are immediately apparent. In models 2 and 6, we see that the Szekeres dipole has very little radial dependence in the range tested. In model 3, shells outside the Szekeres anisotropy spike see virtually no Szekeres dipole at all, while interior shells see a significant amount. The shell in the middle of the spike sees a Szekeres dipole roughly (but not exactly) half the magnitude (and half the $\theta_0$ deviation from $\pi/2$) of the interior shells. In general, as we traverse through shells which have Szekeres anisotropies, the magnitude of the Szekeres dipole decreases and the angle decreases towards $\pi/2$. Furthermore, models in which the Szekeres anisotropies occur at higher $r$ (e.g.\ model 5 compared to 4, or 6 compared to 3) appear to generate Szekeres dipoles with angles deviating less from $\pi/2$, compared to the differences in the Szekeres dipole magnitudes. 

The picture appears to be that the behavior of the Szekeres functions at $r$ values lower than that of the observer has much less effect on the dipole than the behavior at higher $r$ values. As we move outwards from $r_1$ to $r_2$, the portion of the Szekeres functions between $r_1$ and $r_2$ loses its impact. At least for observers reasonably close to the origin (on the order of a few hundred Mpc or less), the effects of the Szekeres functions in the interior are virtually nonexistent.

As found by \cite{Alnes2}, the LTB dipole component increases approximately linearly with $r$ near the origin. However, the Szekeres modifications shift the region of interest to higher $r$ values. We find that a cubic fit matches the data to within $3 \times 10^{-4}$ mK for $r\leq 400$.

\subsection{Size of ``allowed'' region}

For the model to be consistent with observations, one requirement is that the CMB dipole does not greatly exceed the actual observed dipole. A number of authors have found that in LTB models large enough to explain the observed acceleration, this is only true within a very small region near the center; everywhere else, the dipole is much larger \cite{Foreman,Alnes,Alnes2,Yoo2,GBH2,Zibin,Grande,Biswas}. It would therefore seem highly improbable that we would find ourselves in such a specific region where the dipole is relatively small. We wish to repeat this calculation in our Szekeres test models, to see the size and shape of this ``allowed'' region and determine whether there is any measurable advantage over LTB.

We will use our fits for the magnitudes and directions of the Szekeres and LTB dipole components to find the region where the total dipole is less than the 3.35 mK dipole observed by COBE \cite{COBE}. (A more complete calculation would incorporate an additional stochastic dipole component arising from peculiar velocities, but a rigorous calculation of this sort would require knowledge of the evolution of perturbations in a Szekeres model, so we will leave this to future work.\footnote{Since the region of interest is not necessarily near the coordinate origin, the shear may be significant, so we cannot assume that perturbations evolve the same way as in FLRW, as done for LTB in \cite{Foreman}.}) The dipole is only low where the LTB dipole and Szekeres dipole nearly cancel; it must therefore be centered around a point on the shell where the Szekeres dipole magnitude lines intersect the LTB dipole line in Fig.\ \ref{fig:szekdipoles}. Once we have calculated the boundaries of this ``allowed'' region, we will numerically integrate over it to find the mass and volume contained within it.

\begin{figure}[tbp]
	\begin{center}
(a) \\	\includegraphics[height=6.4cm]{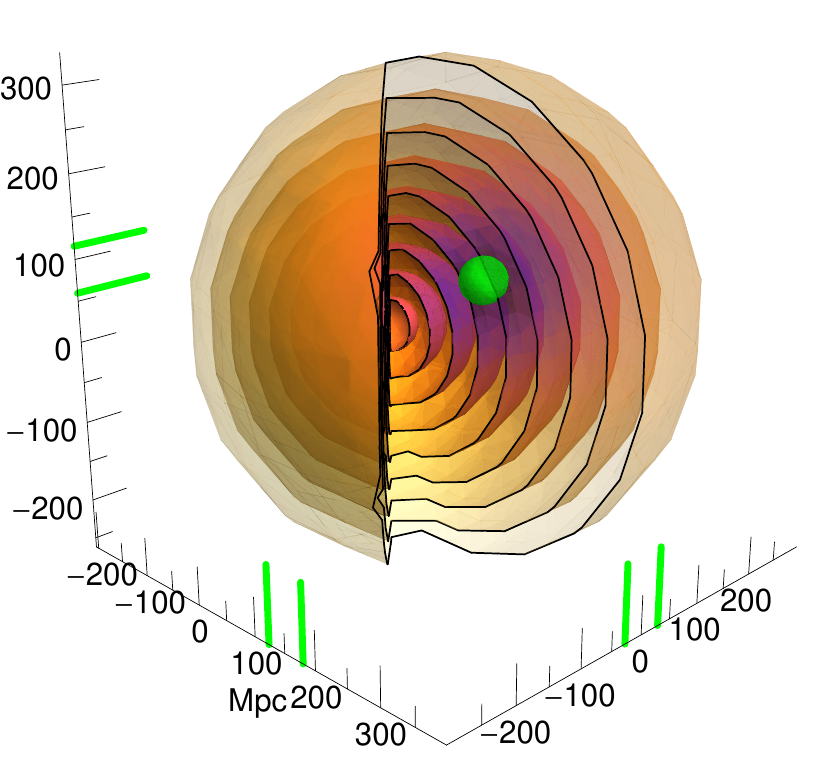} \\
(b) \\	\includegraphics[height=6.4cm]{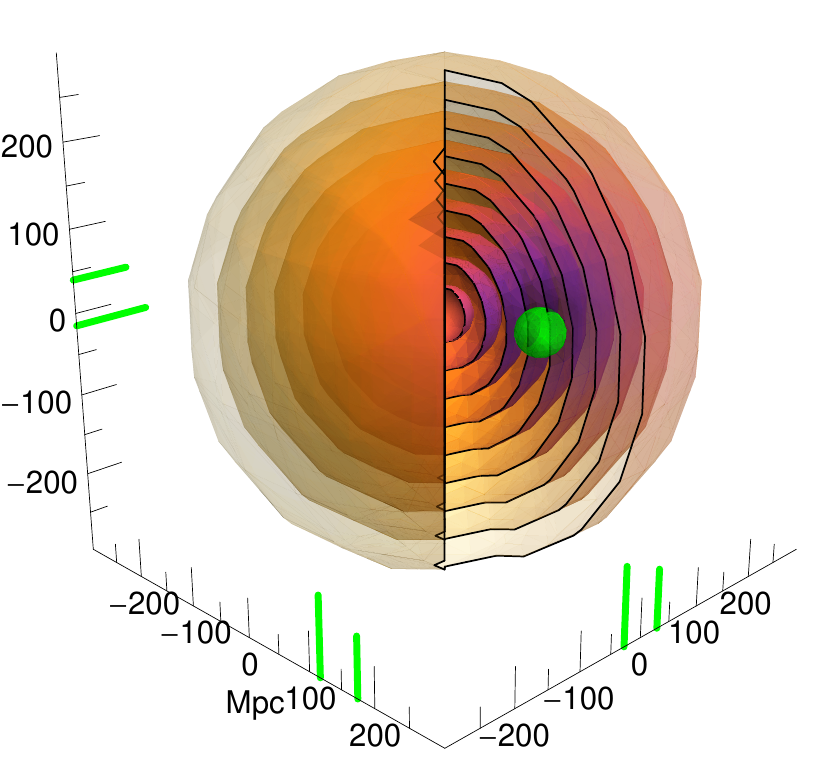} \\
(c) \\	\includegraphics[height=6.4cm]{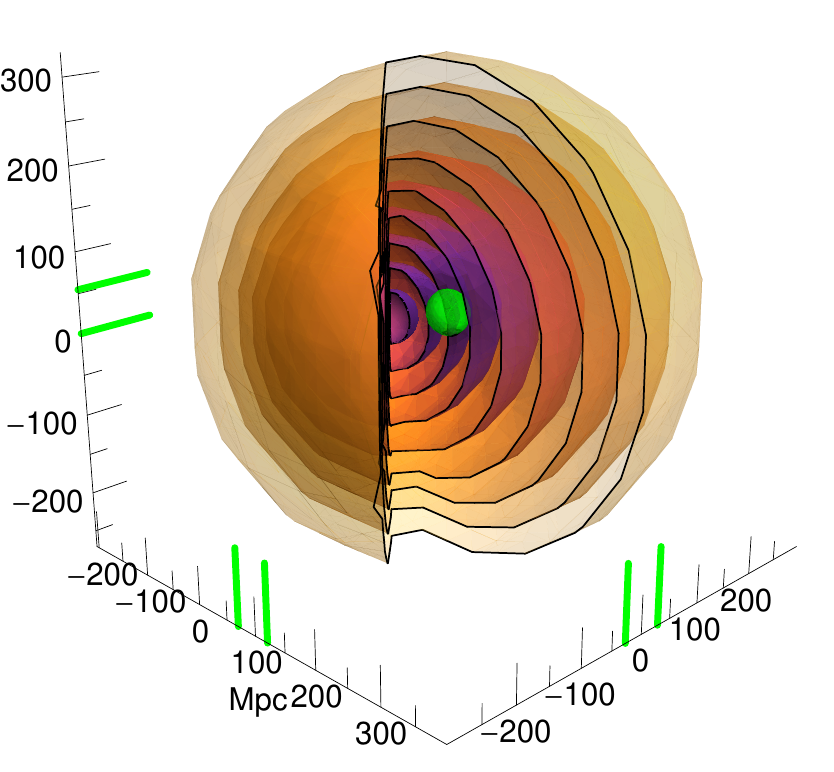}
	\caption{The total dipole magnitudes across models 1 (a), 2 (b), and 3 (c). The region where the dipole is less than the actual observed dipole is shown as a green sphere. Its range is also marked in green on the axes. Shells of constant $r$ are also shown, in increments of 33 Mpc, colored according to the magnitude of the total dipole, with lighter being larger. (Color online)}
	\label{fig:allowedregions}
	\end{center}
\end{figure}

Our results are summarized in Table \ref{tab:allowedregions}. Figure \ref{fig:allowedregions} shows this region visually for each of the three models. We see that they are still small, roughly spherical regions (even when they reside in a region of significant shell shifting and twisting), though they are displaced away from the coordinate origin. They are often larger than in the base LTB model, but still small compared to the size of the void, by a factor on the order of $10^{-6}$.

\begin{table}[tbp]
	\begin{center}
		\begin{tabular*}{8.6cm}{@{\extracolsep{\fill}}ccc}
			\hline \hline
			Model	& \quad	$V_i/V_{\mathrm{LTB}}$ \quad	& \quad	$m_i/m_{\mathrm{LTB}}$ \quad	\\
			\hline
			1 		&	1.14					&	1.26					\\
			2 		&	1.14					&	1.27					\\
			3 		&	0.93					&	0.99					\\
			4 		&	1.18					&	1.35					\\
			5 		&	1.30					&	1.58					\\
			6 		&	1.04					&	1.05					\\
			\hline \hline
		\end{tabular*}
	\caption{Volumes and masses of the ``allowed'' region in each of the six models, compared to that of the base LTB model.}
	\label{tab:allowedregions}
	\end{center}
\end{table}

The mass is the more relevant quantity, since it determines the number of ``allowed'' galaxies. And we should expect that the ``allowed'' mass is in general larger in Szekeres models than in LTB, because the Szekeres anisotropy shifts the ``allowed'' region away from the center of the void. This means it is in a higher density region, with more galaxies where we may find ourselves located. It seems that removing the spherical symmetry of LTB does tend to somewhat alleviate the need for fine-tuning of the observer's location, but not by nearly enough to fix the problem entirely.

\subsection{Higher order multipoles}

Because a complete CMB map is far more computationally intensive, we have fewer data points for the higher order multipoles at this time, so our analysis is limited. We leave a more thorough analysis for future work, and present our preliminary results here.

\begin{figure}[tbp]
	\begin{center}
	(a) \\ \includegraphics[height=4.45cm]{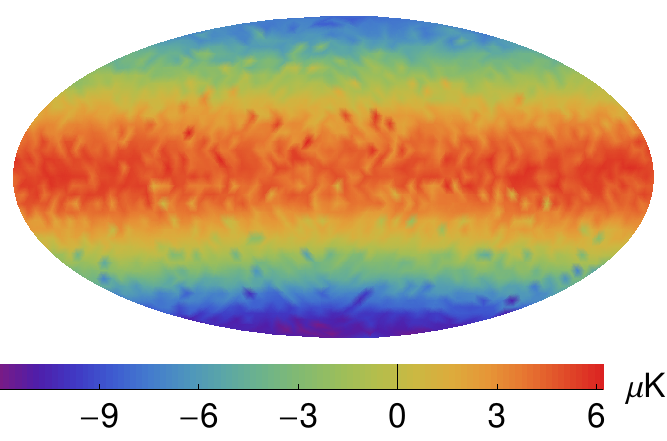} \\
	(b) \\ \includegraphics[height=4.45cm]{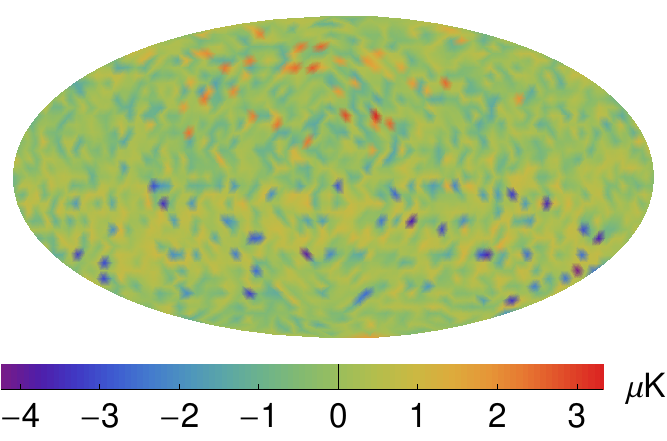} \\
	(c) \\ \includegraphics[height=4.45cm]{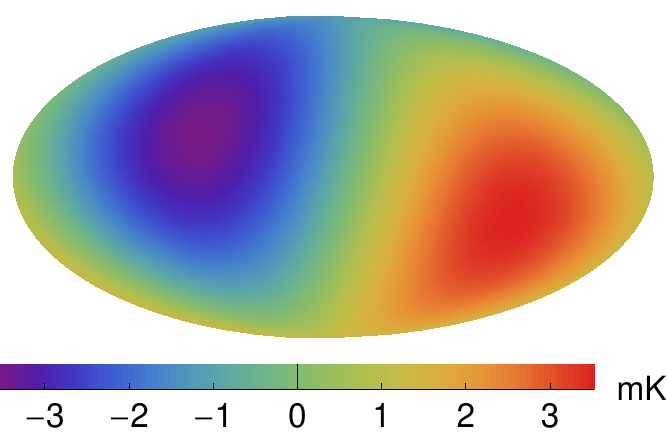} \\
	(d) \\ \includegraphics[height=4.45cm]{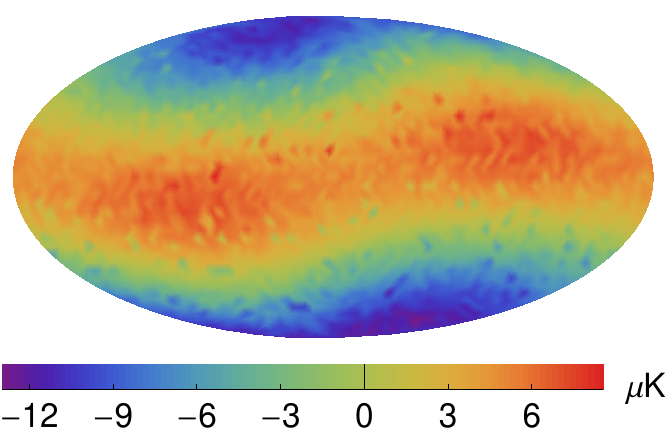} \\
	\caption{The full cmb sky induced by the Szekeres void of model 2. Maps are oriented such that the $z$ axis (the top of the map) points in the model's radial direction, and the center of the map points in the model's $\hat{\theta}$ direction. (a): raw CMB sky map for an observer near the center of the low-dipole region, with only the uniform 2.725 K monopole removed; (b): same, but with dipole and quadrupole removed, showing that no higher moments are visible above the noise. (c): raw map for an observer at a random point near the edge of the low-dipole region; (d): same, but with dipole removed. (Color online)}
	\label{fig:cmb2}
	\end{center}
\end{figure}

We performed the calculation for an observer at the point of zero total dipole in each of the six models. We found that model 2 has a significant quadrupole at this location---about $5 \times 10^{-6}$, compared to the real observed anisotropies of the order $10^{-5}$ \cite{Alnes2}---and a very small octupole, on the order of $10^{-7}$. Figure \ref{fig:cmb2} shows the cmb map at this point, as well as at a random point near the edge of the ``allowed'' region. In the other models, the quadrupole and octupole at the null-dipole point are below the level of the random noise from numerical errors, and are therefore not measurable.

It is clearly not a fluke that the quadrupole vanishes at the null-dipole, since it happened in five very different models, but it does not appear to be a general rule for Szekeres models either, as seen in model 2. The distinguishing feature of model 2 is that the Szekeres anisotropies cover a broad range, reaching a very high $r$ value. We may hypothesize that this is the reason for the difference in behavior---why the quadrupole is nonzero at the point where the dipole vanishes. Model 6 also has anisotropy at high $r$, but Fig.\ \ref{fig:szekdipoles} shows that its total impact on the dipole is relatively small, and it stands to reason that its effect on the quadrupole might be small as well---too small to push it significantly away from zero at the null-dipole point. It seems that a broader range is necessary to visibly affect the quadrupole separately from the dipole.

To test this hypothesis, we created a seventh model, with $C = 0.945$, $r_i = 1500$, and $r_f = 2500$. This is similar to model 6 in that the Szekeres functions only act in the outer regions of the void, but the broader range gives the Szekeres dipoles greater strength. In fact, the magnitude of the Szekeres dipole seen in the inner regions ($r < 300 \text{ Mpc}$) is within 3\% of what is seen in model 2, with the direction the same to within 0.01 radians. The quadrupole at the null-dipole point in model 7, however, is double what it is in model 2---a full $10^{-5}$, comparable to observations. The octupole is still only on the order of $10^{-7}$, though. Comparing models 2 and 7 seems to confirm that, given equal Szekeres dipole strength, the model with Szekeres functions weighted at higher $r$ values will have a larger CMB quadrupole at the null-dipole point.

\begin{figure}[tbp]
	\begin{center}
	\includegraphics[width=8.5cm]{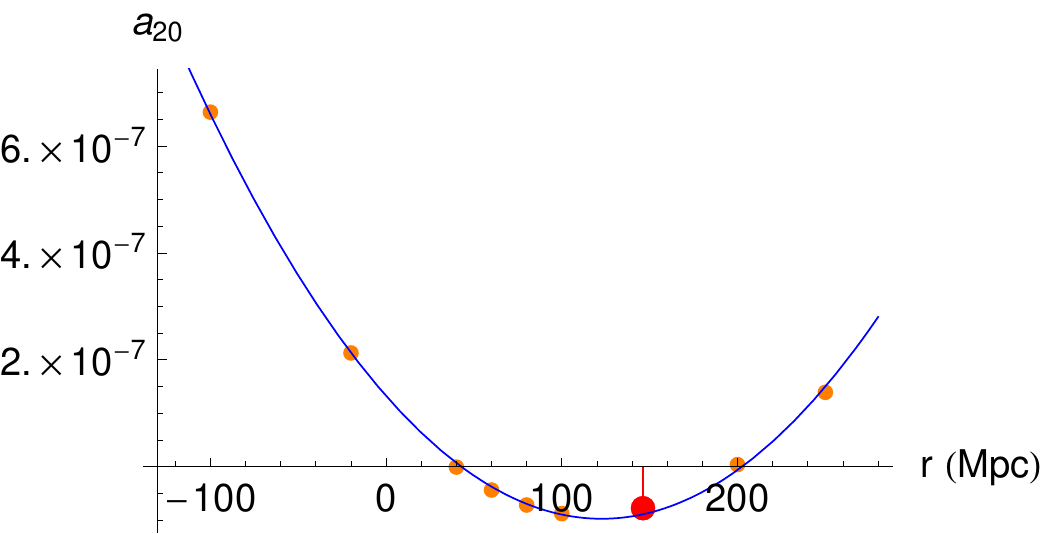}
	\caption{The primary quadrupole coefficient $a_{20}$ at several points along the radial line containing the null-dipole point in model 2, and a quadratic fitting curve. Negative $r$ values simply refer to points on the opposite side of the origin. The larger red dot indicates the null-dipole point. (Color online)}
	\label{fig:quadrupoles}
	\end{center}
\end{figure}

\begin{figure}[tbp]
	\begin{center}
(a) \\	\includegraphics[width=8.5cm]{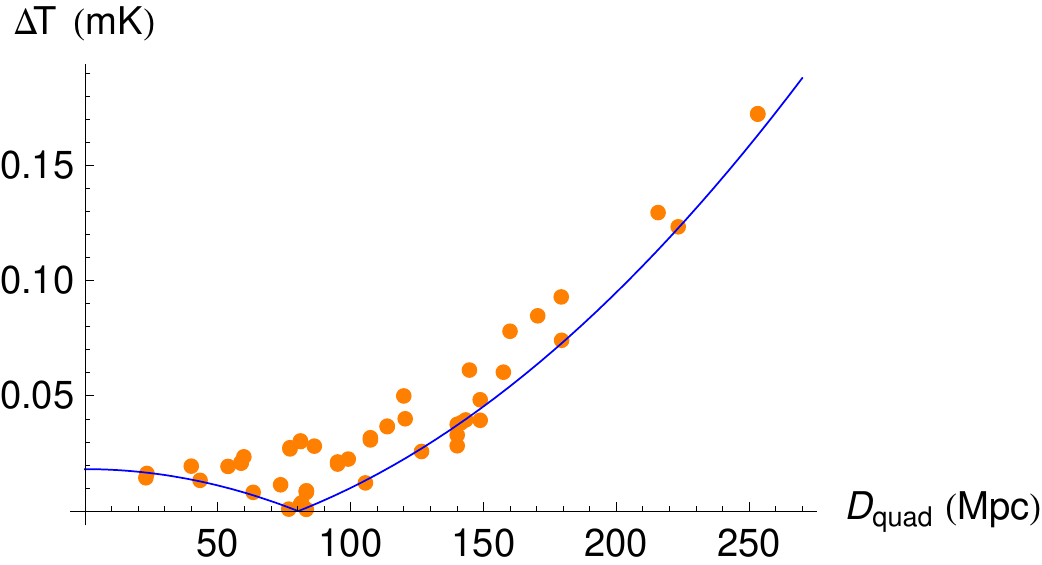} \\
(b) \\  \includegraphics[width=8.5cm]{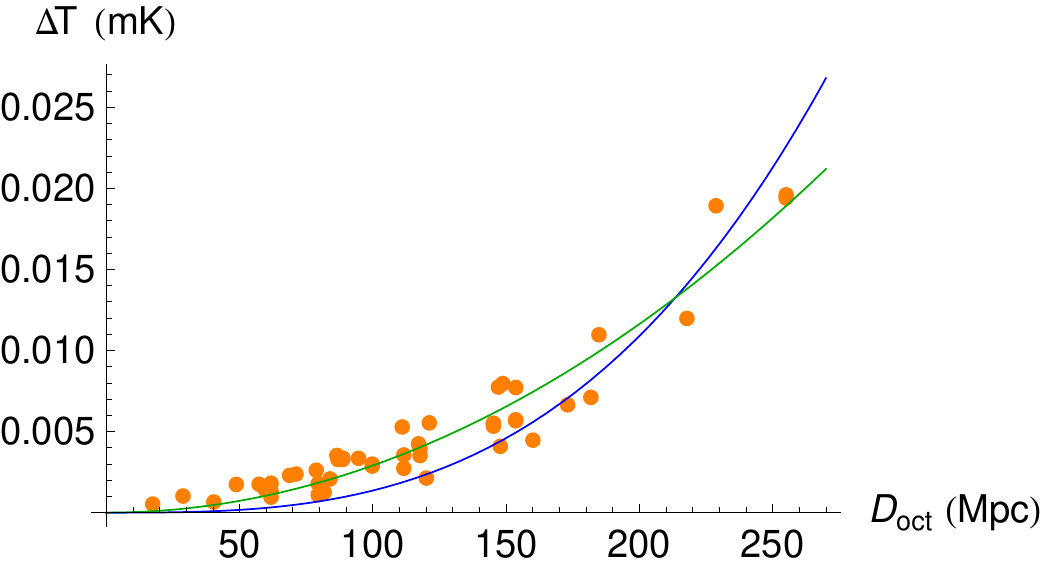}
	\caption{(a) The quadrupoles at all tested points in model 2, in terms of $\Delta T$, as a function of the distance from the center of the fit shown in Fig.\ \ref{fig:quadrupoles}. The blue curve is a simple extrapolation of the fit from Fig.\ \ref{fig:quadrupoles}. (b) The octupoles at all tested points, as a function of the distance from the center of a cubic fit on the line containing the null-dipole point. Both quadratic (green) and cubic (blue) fitting curves are shown for comparison. (Color online)}
	\label{fig:allquadoct}
	\end{center}
\end{figure}

To better understand the more general behavior of the quadrupole and octupole, we gathered data at a number of different points in model 2 (with only 6 degree resolution for faster computations). Along the radial line passing through the null-dipole point, we found that the quadrupole is dominated by $a_{20}$, which follows a simple quadratic curve, as shown in Fig.\ \ref{fig:quadrupoles}. This parabola is centered neither at the origin nor at the null-dipole point, and its minimum dips significantly into the negative. The total quadrupole magnitude thus hits zero at two points on this line, with a hill in between (where the null-dipole point falls). Off of this line, the quadrupole displays more complex behavior, which we do not yet have enough data points to fully describe or explain. Figure \ref{fig:allquadoct} summarizes both the quadrupole and octupole data. The quadrupoles seem to roughly follow a quadratic trend, consistent with what Alnes found for LTB models \cite{Alnes2}, but it is clearly not an exact fit. For the octupoles, it is even less clear that a cubic fit is accurate.

Finally, a test of the CMB at $r = 300$ in model 3, compared with a similar test in the corresponding LTB model, revealed that Szekeres behavior on shells interior to the observer's shell has negligible effects on the entire CMB, not just the dipole. The differences between the two maps are on the level of 1 $\mu$K (a tenth the strength of even the octupole), and appear to follow a random noise pattern across the entire sky; we can thus attribute these small differences to numerical error.

\section{\label{sec:conclusion}Conclusions}

In this paper, we have studied the CMB dipole seen by observers in a Szekeres model. We have established a procedure for calculating dipoles at general locations, and we have shown that they follow a simple, consistent pattern. While the models tested show little quantitative advantage over LTB in terms of the size of the region allowed by dipole observations, Szekeres models do offer greater freedom in where this region is located. We are no longer required to be at the center of the void, where the density is low and anisotropies are only significant at the dipole level.

We have found that the CMB quadrupole seen by observers in the low-dipole region is not always as small as in the corresponding LTB model, and significant compared to the quadrupole seen in the WMAP data. The octupole is still small in this region in all the models tested, but it is possible that a more extreme Szekeres model would amplify that mode as well. There is then some hope that a Szekeres model may offer a possible explanation for the WMAP quadrupole and octupole anomalies.

Of the four shortcomings of LTB listed in section \ref{sec:intro}, it appears that Szekeres models offer improvements on one and a half. The region allowed by the dipole requirements is still small, so there is still a need for fine-tuning of the observer's location, but this region is not necessarily ``special'' in other ways, as it is in LTB void models. That is, LTB void models constrain the observer to a small region that sees a small CMB dipole, and also happens to see a very small quadrupole and octupole, lie near the unique symmetry center of the entire model, and typically be the region of minimum density, whereas a Szekeres void model constrains the observer to a region that is only special in the first of these ways. The strength of the quadrupole and octupole in this region show significant improvement over LTB for some models, but not for others, and it is still unclear whether they can truly match the anomalies seen by WMAP. The kSZ effect, though not calculated here, is expected to still be a problem for Szekeres void models, since the total dipoles still follow a roughly linear trend similar to the LTB model.

It is worth noting that the test models considered here used a homogeneous bang time function, meaning no decaying modes are present. While this is consistent with the standard view of inflation and the early universe, it has been suggested that even slight variations in the bang time could significantly reduce the kSZ effect and allow for very different void profiles. This could be an avenue of future work.

\begin{acknowledgements}
This research was partially supported by the Texas Space Grant Consortium.  We would also like to thank Dr. Mustapha Ishak and Dr. Patrick Greene for helpful comments, and Dr. Edward (Rob) Robinson for the initial suggestion that led to this line of research. Computations were performed with the Wolfram Mathematica 7 software.
\end{acknowledgements}

\appendix

\section{Geodesic equations}
\label{app:geoeq}
Here we write out the full null geodesic equations in the quasispherical Szekeres model, which can also be found in \cite{Nwankwo} or \cite{KrasinskiRedshiftProp}.
\begin{widetext}
\begin{align}
	\label{eq:tgeo}
\frac{\mathrm{d}^2t}{\mathrm{d}s^2} &= -\frac{\Phi,_{tr} - \Phi,_t E,_r/E}{1 - k}(\Phi,_r - \Phi E,_r/E)\left(\frac{\mathrm{d}r}{\mathrm{d}s}\right)^2 - \frac{\Phi\Phi,_t}{E^2}\left[\left(\frac{\mathrm{d}x}{\mathrm{d}s}\right)^2 + \left(\frac{\mathrm{d}y}{\mathrm{d}s}\right)^2\right] \\
	\label{eq:rgeo}
\frac{\mathrm{d}^2r}{\mathrm{d}s^2} &= -2\frac{\Phi,_{tr} - \Phi,_t E,_r/E}{\Phi,_r - \Phi E,_r/E}\frac{\mathrm{d}t}{\mathrm{d}s}\frac{\mathrm{d}r}{\mathrm{d}s} - \left(\frac{\Phi,_{rr} - \Phi E,_{rr}/E}{\Phi,_r - \Phi E,_r/E} - \frac{E,_r}{E} + \frac{1}{2}\frac{k,_r}{1-k}\right)\left(\frac{\mathrm{d}r}{\mathrm{d}s}\right)^2 - 2\frac{\Phi}{E^2} \frac{E,_r E,_x - EE,_{xr}}{\Phi,_r - \Phi E,_r/E}\frac{\mathrm{d}r}{\mathrm{d}s}\frac{\mathrm{d}x}{\mathrm{d}s} \nonumber \\
  &\quad  - 2\frac{\Phi}{E^2} \frac{E,_r E,_y - EE,_{yr}}{\Phi,_r - \Phi E,_r/E}\frac{\mathrm{d}r}{\mathrm{d}s}\frac{\mathrm{d}y}{\mathrm{d}s} + \frac{\Phi}{E^2} \frac{1-k}{\Phi,_r - \Phi E,_r/E}\left[\left(\frac{\mathrm{d}x}{\mathrm{d}s}\right)^2 + \left(\frac{\mathrm{d}y}{\mathrm{d}s}\right)^2\right] \\
	\label{eq:xgeo}
 \frac{\mathrm{d}^2x}{\mathrm{d}s^2} &= -2 \frac{\Phi,_t}{\Phi}\frac{\mathrm{d}t}{\mathrm{d}s}\frac{\mathrm{d}x}{\mathrm{d}s} + \frac{1}{\Phi} \frac{\Phi,_r - \Phi E,_r/E}{1-k}(E,_r E,_x - EE,_{xr})\left(\frac{\mathrm{d}r}{\mathrm{d}s}\right)^2 - \frac{2}{\Phi}\left(\Phi,_r - \Phi\frac{E,_r}{E}\right)\frac{\mathrm{d}r}{\mathrm{d}s}\frac{\mathrm{d}x}{\mathrm{d}s} + \frac{E,_x}{E}\left(\frac{\mathrm{d}x}{\mathrm{d}s}\right)^2 \nonumber \\
 &\quad + 2\frac{E,_y}{E}\frac{\mathrm{d}x}{\mathrm{d}s}\frac{\mathrm{d}y}{\mathrm{d}s} - \frac{E,_x}{E}\left(\frac{\mathrm{d}y}{\mathrm{d}s}\right)^2 \\
	\label{eq:ygeo}
\frac{\mathrm{d}^2y}{\mathrm{d}s^2} &= -2 \frac{\Phi,_t}{\Phi}\frac{\mathrm{d}t}{\mathrm{d}s}\frac{\mathrm{d}y}{\mathrm{d}s} + \frac{1}{\Phi} \frac{\Phi,_r - \Phi E,_r/E}{1-k}(E,_r E,_y - EE,_{yr})\left(\frac{\mathrm{d}r}{\mathrm{d}s}\right)^2 - \frac{2}{\Phi}\left(\Phi,_r - \Phi\frac{E,_r}{E}\right)\frac{\mathrm{d}r}{\mathrm{d}s}\frac{\mathrm{d}y}{\mathrm{d}s} - \frac{E,_y}{E}\left(\frac{\mathrm{d}x}{\mathrm{d}s}\right)^2 \nonumber \\
 &\quad + 2\frac{E,_x}{E}\frac{\mathrm{d}x}{\mathrm{d}s}\frac{\mathrm{d}y}{\mathrm{d}s} + \frac{E,_y}{E}\left(\frac{\mathrm{d}y}{\mathrm{d}s}\right)^2
\end{align}
\end{widetext}

The temperature of the LSS along any geodesic depends on the redshift. This is easy to calculate from the definition of redshift \cite{Ellis}:
\begin{align}
1+z = \frac{(k_\alpha u^\alpha)_s}{(k_\alpha u^\alpha)_o},
\end{align}
where subscripts $s$ and $o$ denote source and observer respectively, $k^\alpha = \mathrm{d}x^\alpha / \mathrm{d}\lambda$, $u$ is the four-velocity of the source or observer, defined to be $(1,0,0,0)$ because the matter is comoving. We can normalize the null geodesic tangent vector at the observer so that $k^t_o = -1$, so we are left with simply 
\begin{align}
1+z = -k^t_s.
\end{align}

\section{Choosing directions}
\label{app:choosingdirections}

We will label the three geodesics with subscripts $1$, $2$, and $3$. Greek indices will refer to spacetime dimensions, while latin indices will refer only to spatial dimensions $r$, $x$, and $y$.

Each geodesic is defined by three initial tangent vector components, $k^r$, $k^x$, and $k^y$ ($k^t$ being determined by the null condition), but there is a degree of freedom in the scale of the affine parameter that allows us to remove a constant factor from each component (a geodesic with the entire tangent vector doubled is still the same geodesic). Since we need three pairs of opposite geodesics, we necessarily have three with positive $k^r$ and three with negative $k^r$ (assuming none are 0). We can therefore decide that we will focus on the ones with positive $k^r$, and scale them so that they all in fact share the same $k^r$, which we choose arbitrarily. This leaves two degrees of freedom for the choice of direction for each geodesic, so we need six equations to fix them.

The only strict requirement is mutual spatial orthogonality. With three geodesics with initial tangent vectors $k_1^\alpha$, $k_2^\alpha$, and $k_3^\alpha$, we have three spatial orthogonality equations:
\begin{subequations}
\label{eq:orthogonality}
\begin{align}
g_{ij}k_1^i k_2^j &= 0 \\
g_{ij}k_2^i k_3^j &= 0 \\
g_{ij}k_3^i k_1^j &= 0.
\end{align}
\end{subequations}
We still need three more, which we can choose more or less arbitrarily. 

We wish to keep the geodesics away from the axis, where $x$ and $y$ go to infinity or zero, since the numerical integration of the geodesic equations loses precision here. To do this, we try to maximize the quantity $\left| \cos \phi k^y  - \sin \phi k^x \right|$.
Due to the orthogonality, the magnitude of this quantity for one geodesic can only be increased at the expense of another. We therefore choose to require that all three have the same magnitude. To satisfy orthogonality, we will need to have two with the same sign and one with the opposite. This gives us two equations,
\begin{subequations}
\begin{align}
\label{eq:phidirection1}
\cos \phi k_1^y - \sin \phi k_1^x &= 
\cos \phi k_2^y - \sin \phi k_2^x, \\
\label{eq:phidirection2}
\cos \phi k_1^y - \sin \phi k_1^x &= -\cos \phi k_3^y + \sin \phi k_3^x.
\end{align}
\end{subequations}
For the final equation, we choose 
\begin{align}
\label{eq:thetadirection}
&\cos \phi \left(k_1^x + k_2^x\right)
 + \sin \phi \left(k_1^y + k_2^y\right)  = 0. 
\end{align}

These six equations have two distinct solutions, corresponding roughly to (1) geodesics going right, up-left, and down-left, and (2) geodesics going left, up-right, and down-right. Which we choose is not important. For consistency, we will simply require $\left(\mathrm{d}\phi /\mathrm{d}s\right)_3 < 0$, corresponding to solution (2). A basic picture of the geodesics generated by these methods is shown in Fig.\ \ref{fig:6geo}.

\section{Confirmation of 6-geodesic dipole equation}
\label{app:6geo}

To see that Eq.\ (\ref{eq:6dipole}) indeed gives the correct dipole, and to estimate the error caused by the quadrupole (expected to typically be the next largest multipole moment), we can expand the CMB temperature into spherical harmonics to the second degree.
\begin{align}
\label{eq:CMBsphericalharmonics}
T &= T_0 \left(1 + \sum_{m=-1}^1 {a_{1m}Y_{1m}} + \sum_{m=-2}^2 {a_{2m}Y_{2m}}\right)
\end{align}
For simplicity, we can orient our sky so that geodesics 1, 3, and 5 go in the directions $(\theta ,\phi ) = (0,0), (\pi/2, 0), (\pi/2, \pi/2)$, and 2, 4, and 6 in the opposite directions. The terms in the numerator of (\ref{eq:6dipole}) have no net contribution from the quadrupole terms, since $Y_{2m}(\pi - \theta, \phi + \pi) = Y_{2m}(\theta ,\phi )$. In the denominator, the dipole terms cancel out in a similar fashion. We then find 
\begin{widetext}
\begin{align}
\label{eq:6dipolespherharm}
v &= \frac{\sqrt{\frac{3}{\pi} \left|a_{10}\right|^2 + \frac{3}{2\pi} \left|a_{1,-1} - a_{11}\right|^2 + \frac{3}{2\pi} \left|-i a_{1,-1} -i a_{11}\right|^2}}{2 + \frac{1}{3} \sqrt{\frac{5}{\pi}} a_{20} + \frac{1}{3} \sqrt{\frac{15}{2\pi}} (a_{22} - a_{2,-2}) - \frac{1}{6} \sqrt{\frac{5}{\pi}} a_{20} - \frac{1}{3} \sqrt{\frac{15}{2\pi}} (a_{22} - a_{2,-2}) - \frac{1}{6} \sqrt{\frac{5}{\pi}} a_{20}} \notag \\
 &= \sqrt{\frac{3}{4\pi}} \sum_{m=-1}^1 \left|a_{1m}\right|^2.
\end{align}
\end{widetext}
So we see that this prescription gives the correct dipole, and the quadrupole introduces no error. Additional contributions come only from the octupole terms and higher.


\begin{thebibliography}{}

\bibitem{Riess} A. G. Riess {\itshape et al.}, Astron. J. {\bfseries 116}, 1009 (1998) [arXiv:astro-ph/9805201]

\bibitem{Larson} D. Larson {\itshape et al.}, Astrophys. J. Suppl. {\bfseries 192}, 16 (2011) [arXiv:1001.4635]

\bibitem{Percival} W. Percival {\itshape et al.}, MNRAS {\bfseries 401}, 2148 (2010)

\bibitem{Durrer} R. Durrer, Phil. Trans. R. Soc. A {\bfseries 369}, 1 (2011) [arXiv:1103.5331].

\bibitem{Buchert} T. Buchert, Gen. Rel. Grav. {\bfseries 40}, 467 (2008) [arXiv:0707.2153]

\bibitem{Wiltshire} D. L. Wiltshire, New J. Phys. {\bfseries 9}, 377 (2007) [arXiv:gr-qc/0702082]

\bibitem{Alnes} H. Alnes, M. Amarzguioui, and \O . Gr\o n, Phys. Rev. D {\bfseries 73}, 083519 (2006) [arXiv:astro-ph/0512006].

\bibitem{Celerier} M.-N. C\'eleri\'er, New Advances in Physics {\bfseries 1}, 29 (2007) [arXiv:astro-ph/0702416].

\bibitem{Clarkson} C. Clarkson and R. Maartens, Class. Quantum Grav. {\bfseries 27}, 124008 (2010) [arXiv:1005.2165].

\bibitem{GBH} J. Garcia-Bellido and T. Haugb\o lle, JCAP {\bfseries 04}, 003 (2008) [arXiv:0802.1523].

\bibitem{GBH2} J. Garcia-Bellido and T. Haugb\o lle, JCAP {\bfseries 09}, 016 (2008) [arXiv:0807.1326].

\bibitem{Alnes2} H. Alnes and M. Amarzguioui, Phys. Rev. D {\bfseries 74}, 103520 (2006) [arXiv:astro-ph/0607334].

\bibitem{Foreman} S. Foreman, A. Moss, J. P. Zibin, and D. Scott, Phys. Rev. D {\bfseries 82}, 103532 (2010) [arXiv:1009.0273].

\bibitem{Zibin} J. P. Zibin and A. Moss, Class. Quant. Grav {\bfseries 28}, 164005 (2011) [arXiv:1105.0909].

\bibitem{Biswas} T. Biswas, A. Notari, and W. Valkenburg, JCAP {\bfseries 11}, 030 (2010) [arXiv:1007.3065].

\bibitem{Celerier2} M.-N. C\'eleri\'er, Astron. Astrophys. {\bfseries 518}, A21 (2010) [arXiv:0906.0905].

\bibitem{Romano} A. E. Romano, Phys. Rev. D {\bfseries 75}, 043509 (2007) [arXiv:astro-ph/0612002].

\bibitem{Zibin2} J. P. Zibin, Phys. Rev. D {\bfseries 78}, 043504 (2008) [arXiv:astro-ph/0612002].

\bibitem{Yoo} C.-M. Yoo, T. Kai, K.-i. Nakao, Phys. Rev. D {\bfseries 83}, 043527 (2011) [arXiv:1010.0091].

\bibitem{Bolejko2} K. Bolejko, M.-N. C\'eleri\'er, and A. Krasi\'nski, Class. Quantum Grav. {\bfseries 28}, 164002 (2011) [arXiv:1102.1449].

\bibitem{BolejkoCopernican} K. Bolejko and J. S. B. Wyithe, JCAP {\bfseries 02}, 020 (2009) [arXiv:0807.2891].

\bibitem{ClarksonCopernican} C. Clarkson, B. A. Bassett, and T. H.-C. Lu, Phys. Rev. Lett. {\bfseries 101}, 011301 (2008) [arXiv:0712.3457].

\bibitem{Caldwell} R. R. Caldwell and A. Stebbins, Phys. Rev. Lett. {\bfseries 100}, 191302 (2008) [arXiv:0711.3459].

\bibitem{Valkenburg} W. Valkenburg, V. Marra, and C. Clarkson, (2012) [arXiv:1209.4078].

\bibitem{Zhang} P. Zhang and A. Stebbins, Phys. Rev. Lett. {\bfseries 107}, 041301 (2011) [arXiv:1009.3967].

\bibitem{Yoo2} C.-M. Yoo, K.-i. Nakao, and M. Sasaki, JCAP {\bfseries 10}, 011 (2010) [arXiv:1008.0469].

\bibitem{Tegmark} M. Tegmark, A. de Oliveira-Costa, and A.J.S. Hamilton, Phys. Rev. D {\bfseries 68}, 123523 (2003).

\bibitem{Bennett} C. L. Bennett {\itshape et al.}, ApJS {\bfseries 192}, 17 (2011).

\bibitem{Szekeres} P. Szekeres, Phys. Rev. D {\bfseries 12}, 2941 (1975); Commun. Math. Phys. {\bfseries 41}, 55 (1975).

\bibitem{Ishak} M. Ishak {\itshape et al.}, Phys. Rev. D {\bfseries 78}, 123531 (2008) [arXiv:0708.2948].

\bibitem{BolejkoCoarse} K. Bolejko and R. A. Sussman, Phys. Lett. B {\bfseries 4}, 265 (2010) [arXiv:1008.3420].

\bibitem{Bonnor} W. B. Bonnor, A. H. Sulaiman, and N. Tomimura, Gen. Relativ. Gravit. {\bfseries 8}, 549 (1977).

\bibitem{Nwankwo} A. Nwankwo, M. Ishak, and J. Thompson, JCAP {\bfseries 05}, 028 (2011) [arXiv:1005.2989]. 

\bibitem{KrasinskiRedshiftProp} A. Krasi\'nski and K. Bolejko, Phys. Rev. D {\bfseries 83}, 083503 (2011) [arXiv:1007.2083].

\bibitem{Grande} J. Grande and L. Perivolaropoulos, Phys. Rev. D {\bfseries 84}, 023514 (2011) [arXiv:1103.4143].

\bibitem{COBE} C. L. Bennett {\itshape et al.}, Astrophys. J. {\bfseries 464}, L1 (1996), [arXiv:astro-ph/9601067].

\bibitem{Ellis} G. F. R. Ellis, in {\itshape General Relativity and Cosmology}, ed. R. K. Sachs, Rend. Scuola Int. Fis. Enrico Fermi, XLVII Corso, New York: Academic Press (1971).


\end{thebibliography}
\end{document}